\documentclass{aastex61}
\usepackage{amsmath}

\newcommand{\Tb}{$T_B$}
\newcommand{\alf}{Alfv\'{e}n\ }
\newcommand{\alfic}{Alfv\'{e}nic\ }
\newcommand{\Alfic}{ALFV\'{E}NIC\ }

\newcommand{\B}{$\vec{B}$}
\newcommand{\n}{$\overline{n}$}
\newcommand{\omgp}{$\omega_p$}

\newcommand{\dn}{$\delta N/N$}

\newcommand{\sigsrc}{$\sigma_{source}$}
\usepackage{amssymb}
\usepackage{xcolor,comment}
\accepted{in ApJ on February 10, 2019}

\shorttitle{Super-\alfic oscillations in Solar type III radio bursts}
\shortauthors{A. Mohan et al.}

\begin{document}

\title{Evidence for Super-\Alfic Oscillations in Solar Type III Radio Burst Sources.}

\correspondingauthor{Atul Mohan}
\email{atul@ncra.tifr.res.in}

\author[0000-0002-1571-7931]{Atul Mohan}
\affil{National Centre for Radio Astrophysics -- Tata Institute of Fundamental Research
Pune 411007, Maharashtra, India.}

\author{Surajit Mondal}
\affil{National Centre for Radio Astrophysics -- Tata Institute of Fundamental Research
Pune 411007, Maharashtra, India.}

\author{Divya Oberoi}
\affil{National Centre for Radio Astrophysics -- Tata Institute of Fundamental Research
Pune 411007, Maharashtra, India.}

\author{Colin J. Lonsdale}
\affil{Massachusetts Institute of Technology -- Haystack Observatory, Westford 01886, Massachusetts, USA.}
 


\begin{abstract}
At the site of their origin, solar meterwave radio bursts contain pristine information about the local coronal magnetic field and plasma parameters. On its way through the turbulent corona, this radiation gets substantially modified due to propagation effects. Effectively disentangling the intrinsic variations in emission from propagation effects has remained a challenge. We demonstrate a way to achieve this, using snapshot spectroscopic imaging study of weak type III bursts using data from the Murchison Widefield Array (MWA). Using this study, we present the first observational evidence for second-scale Quasi-Periodic Oscillations (QPOs) in burst source sizes and orientation with simultaneous QPOs in intensity. The observed oscillations in source sizes are so fast and so large that they require two orders of magnitude larger \alf speed than the typical local value of $0.5\ Mm/s$ at the burst generation heights, if interpreted within a MHD framework. These observations imply the presence of a quasi-periodic regulation mechanism operating at the particle injection site, modulating the geometry of energetic electron beams that generate type III bursts. In addition, we introduce a method to characterize plasma turbulence in mid coronal ranges. We also detect evidence for a systematic drift in the location of the burst sources superposed on the random jitter induced by scattering. We interpret this as the motion of the open flux tube within which the energetic electron beams travel.

\end{abstract}




\section{Introduction} \label{intro}
Type III bursts have been studied at metre-wavelengths over many decades, starting with \citet{wild1950}. These bursts are caused by high energy electron beams accelerated at magnetic field (\B) reconnection sites. These electron beams produce two-stream instabilities along their trajectories which give rise to sites of Langmuir wave turbulence. These sites then produce coherent radio emission at the local plasma frequency (\omgp) and its first harmonic via wave-particle and multi-wave interactions (\citealp{ginzburg1958}; \citealp{Tsytovich69}; \cite{melrose1972}; \citealp{Reid2014}). 
{Earlier studies have shown that for these bursts, quite often only the harmonic emission component is observed (\citealp[e.g.][]{dulk1984}; \citealp{robinson_cairns1994}). 
The dependence of the observed frequency of emission on local plasma density implies that emission at a given frequency must arise from iso-density regions.
This makes spectroscopic observation of type III bursts a unique tool for detailed coronal studies.}

These bursts are known to show many intriguing features like large source sizes, rapid fluctuations in source locations in radio images, Quasi Periodic Oscillations (QPOs) in radio flux density dynamic spectrum etc. (\citealp{wild1970}; \citealp{sheridan1972}; \citealp{dulk1980}; \citealp{raoult1980}; \citealp{Pick1980}; \citealp{dulk1984};  \citealp{Hilaire}). The origin of many of the observed properties of type III burst sources have been a matter of active debate. While some groups argued that the observed source properties are primarily driven by strongly divergent coronal \B , variability of supra-thermal electron beams and other physical conditions at the burst site (\citealp{sheridan1972}; \citealp{Pick1980}; \citealp{raoult1980}; \citealp{Duncan1985}), others argued for the dominance of wave propagation effects - namely ducting, scattering and refraction (\citealp{steinberg1971}; \citealp{robinson_scat1983}; \citealp{bastian1994}; \citealp{Arzner1999};  \citealp{Wu2002}). 
{It is widely believed that the second-scale QPOs observed in type III burst intensity must arise due to a cause intrinsic to the source of emission, rather than radio wave propagation effects through the turbulent corona. 
The underlying reason is that the propagation effects are caused by turbulent density fluctuations in the ambient corona, which are expected to remain statistically steady over timescales of order a few minutes. 
The mechanisms which modulate the observed type III burst intensities can be due to plasma/magnetic field dynamics either locally at the site of generation of metre-wave burst emission or at the particle beam acceleration/injection site situated at much lower coronal heights. \citet{Asch_QPPtheoryRev1987} classify the theories proposed to explain coronal radio pulsations into three classes, namely MHD oscillations, cyclic self-organizing systems and modulation of acceleration. 
The first class of theories explain the modulations in radio emission as a result of MHD oscillatory modes in the flux tube transporting the energetic electrons. These oscillations are modelled as initiated locally either by impulsive disturbances \citep{roberts_fastQPP1984} or due to modulations by an already existing MHD wave \citep{rosenberg1970, Ash2004_SausMode_NthHarmonic_QPO, sharykin2018_LOFAR_dnn_withtypIIIb}. They modulate the local plasma density and in turn emissivity, resulting in QPOs \citep{Asch_observersview2003}. Theories based on cyclic self-organizing systems rely on modulations in wave-wave and wave-particle resonance. The studies of non-linear plasma response to an explosive instability show that the system can go into a quasi-periodic phase where it oscillates between resonant and non-resonant states \citep[e.g][]{Smith1971_2streamInst_inWeakTurb}. 
The third class of theories rely on modulation mechanisms strictly at the particle injection/acceleration sites, as opposed to the sites of origin of burst emission.
The modulation mechanism may involve coupling of MHD eigen modes with reconnection sites or pulsed injection which periodically set up resonance conditions in overlying plasma
\citep{ash94_pulsdAccl}. There is no clear consensus on which class(es) of models can correctly explain most type III QPOs. 
The key bottleneck has been that the ability to image the source of burst emission with sufficiently fine spectral and temporal resolution, essential for making progress on this problem, has not been available.} 

{New generation radio interferometric arrays like the LOw-Frequency ARray (LOFAR; \citealp{vanHaarlem13}), the Karl G. Jansky Very Large Array (JVLA; \citealp{Perley2011}) and the Murchison Widefield Array, (MWA; \citealp{Lonsdale09}; \citealp{Tingay2013}), have facilitated snapshot-spectroscopic (SS) imaging by providing frequency coverage spanning wide bands and sub-second imaging time resolutions. 
There are now many studies in the literature exploiting this  comparatively new capability to study various aspects of solar radio bursts and the ambient corona \cite[e.g.][]{Div11, chen2013_SSimagingJVLA, morosan2014, morosan2017_jburst, Atul17, patrick2018_densmodel_frmtypIII, mcauley2017_QSLwithMWA, cairns2018}. 
SS imaging studies of sources of type III emission can provide a comprehensive spectro-temporal evolutionary picture of all observable properties, namely morphology, intensity and sky location. Using SS imaging with LOFAR, \cite{kontar2017} showed that the size of type III sources display rapid growth rates which can only be explained using radio wave scattering by the medium. This study was based on a very strong event with peak flux density of order $10^5\ SFU$. 

Here we present a detailed SS imaging study of a weak type III burst. 
Such bursts are usually triggered by faint coronal jets or microflares.
Given their low energies, such bursts are not expected to lead to significant structural changes in the pre-existing magnetic, temperature or density fields. A useful analogy is to consider the electron beams arising out of such events as `test particles', which probe a system without altering its state. We study the spectro-temporal evolution of the intensity, size, orientation and sky location of the source of burst emission. 
As alluded to earlier, the emission from the burst sites gets modified due to propagation though the turbulent corona, and it is challenging to disentangle the effects of propagation from the intrinsic variations in the emission.
We introduce a novel approach for separating propagation effects from the intrinsic dynamics of the source of burst emission.}

This paper is organised as follows: Section \ref{obs} describes the event and the observations used for this study. The data analysis procedure and the observed properties of the type III radio source are detailed in Section \ref{analysis}. Section \ref{discussion} presents our interpretation of the various observed trends in properties of the burst source, and Section \ref{conclusion} the conclusions.

\section{Observations}\label{obs}
The data used here come from the MWA observations on 3 November 2014 from 06:12:02-06:16:02 UT. They were taken under the solar observing proposal G0002. These data have a bandwidth of 15.36 MHz, centered at 118.4 MHz, and a time and frequency resolution of 0.5 s and 40 kHz respectively. A total of 10 type III radio bursts, 8 type IV radio bursts and a GOES class C1.4 flare were reported on the day by Space Weather Prediction Center / National Oceanic and Atmospheric Administration (SWPC/NOAA). NOAA active region summary listed 10 active regions on the visible part of the solar disc and the day was characterised as a period of medium activity level\footnote{http://www.solarmonitor.org}. {SWPC reported a group of type III bursts during 06:02 -- 07:22 UT, the period which includes our observations\footnote{ftp://ftp.swpc.noaa.gov/pub/indices/events/README}.} 
  \begin{figure}
      \centering
      \includegraphics[width=18cm,height=6.8cm]{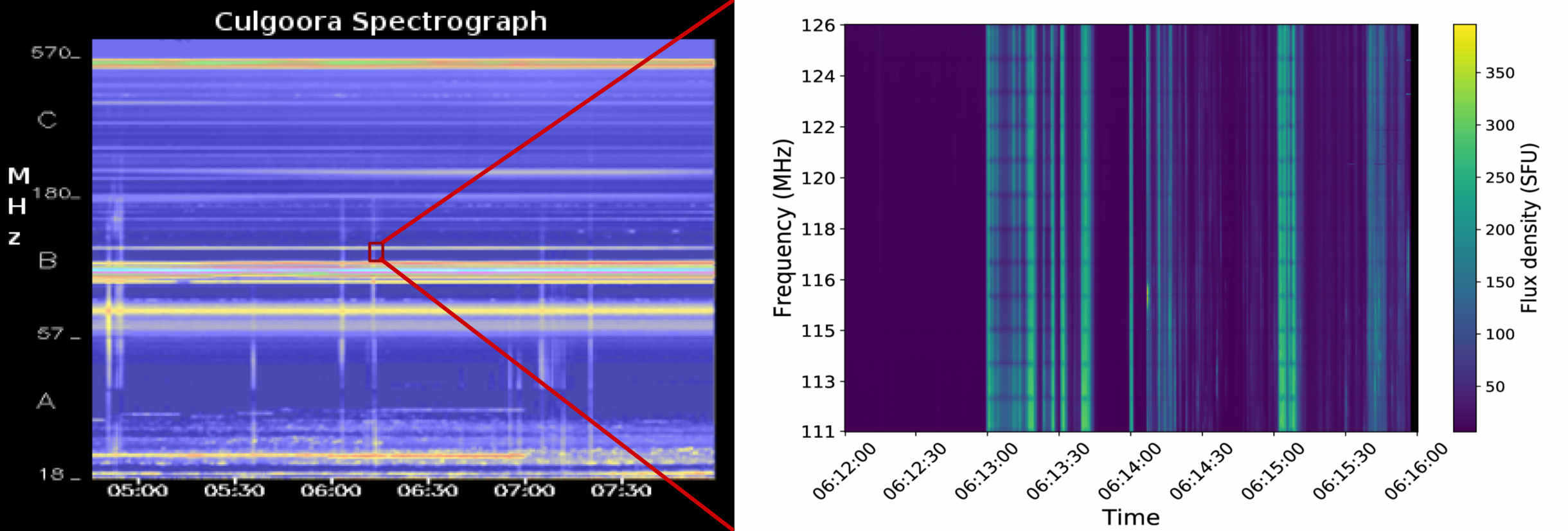}
      \caption{{\it Left: }Dynamic spectrum from the Culgoora spectrograph spanning our observation period and band. The red box marks the spectro-temporal extent of the MWA observation. {\it Right: }The MWA dynamic spectrum showing groups of type III bursts during our observation window.}
      \label{culg+mwa}
  \end{figure}
  \begin{figure}
      \centering
      \includegraphics[width=18cm,height=6.2cm]{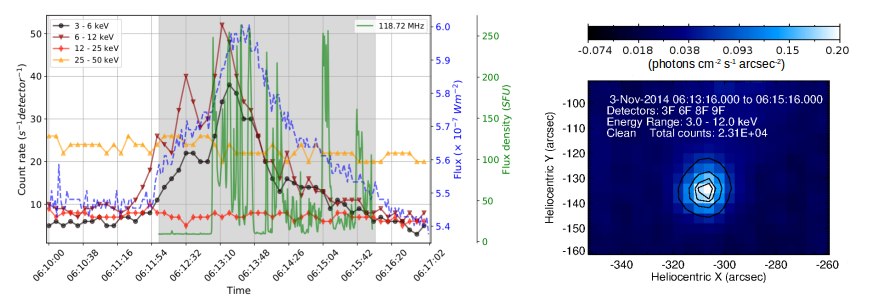}
      \caption{{\it Left: } The RHESSI count rate during the period of our observation. GOES 1 -- 8 \AA\ light curve is shown by dashed lines and the line plot without marker shows 118.72  MHz light curve from the MWA. The shaded region marks the period of the MWA observation.
      {\it Right: }RHESSI image obtained by averaging over 120 s (06:13:16 -- 06:15:16 UT) and 9\ ~keV (3 -- 12\ ~keV). The RHESSI flare region is centred at $(-308^{\prime\prime},-136^{\prime\prime})$ in heliocentric coordinates.}
      \label{compr}
  \end{figure}

The left panel of Fig.\ref{culg+mwa} shows the dynamic spectrum from the Culgoora spectrograph spanning the spectral range from 18--180\ MHz and straddling our observations in time. The multiple near vertical weak emission streaks in the Culgoora dynamic spectra comprise the group of type III bursts. One of these bursts is captured by the MWA observations and the corresponding dynamic spectrum is shown in the right panel of Fig.\ref{culg+mwa}.
What seems like a single streak in the Culgoora dynamic spectrum is revealed to have a more elaborate structure in the MWA data.
The left panel of Fig.\ref{compr} shows full disc light curves for different bands of the Reuven Ramaty High Energy Solar Spectroscopic Imager (RHESSI; \citealp{lin2002}), GOES 1--8 \AA\ emissions and MWA obervations at $118.72\ MHz$.
A small increase in 3 -- 12 keV flux was detected by RHESSI during our observation window\footnote{http://sprg.ssl.berkeley.edu/~tohban/browser/?show=grth+qlpcr+qli02+qlids\& bar=1}. 
The GOES flux also shows a co-temporal increase in the 1--8 \AA\ band. This corresponds to a {B6 GOES class flare (without background subtraction)}\footnote{http://spaceweather.com/glossary/flareclasses.html}. We note that the $118.72\ MHz$ light curve shows bursty emission features which start close to the peak time of the RHESSI and GOES flare, and continue till close to the end of the X-ray flare. We examined the association of the X-Ray and radio events. We made images of the X-Ray flare using the RHESSI analysis routines in SolarSoft Ware (\citealp{freeland1998_SSW}) package based on IDL. {Imaging was done using CLEAN algorithm on the data from detectors identified to be in proper working condition. We identified an X-Ray source at ($-308^{\prime\prime}$,$-136^{\prime\prime}$) (Fig.\ref{compr}, right panel) close to the location of the radio burst source. During the period of the GOES flare another slightly weaker RHESSI source was also present.
It was located at ($150^{\prime\prime}$,$130^{\prime\prime}$) and associated with a different active region.}
\section{Analysis}\label{analysis}
\subsection{Radio-Data analysis}\label{radio_analysis}
The MWA data were imaged every 0.5 seconds at a spectral resolution of 160 kHz across the 15 MHz frequency band for the entire 4 minute duration of observation. Imaging was done using an automated radio interferometric imaging pipeline optimised for MWA like arrays with a strong central condensation of collecting area. Named Automated Imaging Routine for Compact Arrays for Radio Sun (AIRCARS; {\color{blue}Mondal et al. 2019 (Submitted)}), it uses imaging and calibration routines from Common Astronomy Software Applications (CASA; \citealp{casa}). Very briefly, AIRCARS relies on the technique of self-calibration (selfcal; \citealp{pearson1984}; \citealp{cornwell99}) to carefully build up a model for the solar emission and the antenna gains for a chosen time and frequency slice. The initial calibration solutions for the antennas, usually obtained from night time observation of calibrator sources, form the typical starting point for the selfcal cycle. The calibration process starts with using only the densest central part of the array. To this part of the array, the Sun appears as a barely resolved source, leading to a highly over constrained calibration problem. Over a large number of iterations, the calibration solutions and the solar emission model are refined and antennas at increasing distances from the array centre are included. AIRCARS has been demonstrated to routinely achieve imaging dynamic ranges approaching $10^3$, even in the most challenging situations, and those approaching $10^5$ in the most favourable conditions. AIRCARS has provided robust imaging performance over the 80 -- 300 MHz MWA observing range over a diverse range of solar conditions.

A limitation of selfcal based approaches is that they do not preserve the information of the absolute source location. The key implication is that images which have gone through independent selfcal processes are no longer registered on a common astrometric reference frame. However, all the images derived from a given selfcal solution do share a common astrometric reference frame. We overcome the limitations imposed by loss of absolute positions, first by relying only on relative motions in the image plane and then by devising a technique to obtain all images for a given spectral band on a common astrometric reference frame, which is described in detail in Appendix \ref{app1}. The whole data set spanning a time period of 4 minutes and bandwidth of 15 MHz resulted in $\approx 23,000$ images. These were flux calibrated using the technique described in \citet{Div17} and \citet{Atul17}.
\subsection{The Event Analysis}
\begin{figure}[!htb]
    \centering
    \includegraphics[width=14cm,height=6.5cm,clip=]{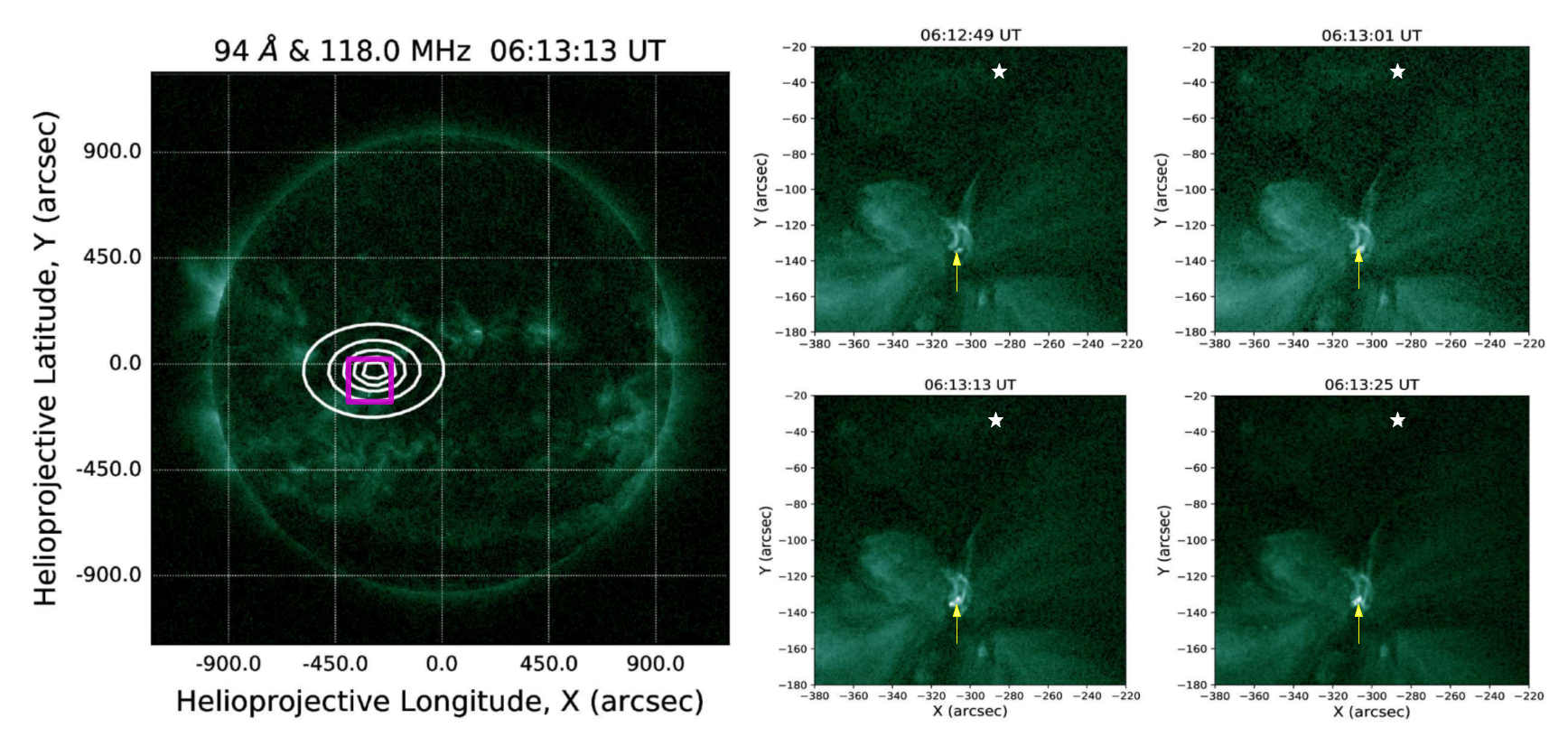}
    \caption{The full disk figure in the left panel shows an AIA 94 \AA\ image with MWA 118 MHz \Tb\ contours overlaid during the mid-time of the first group of bursts. The contours are drawn at 60\%, 80\%, 90\%, 95\% and 98\% of the peak \Tb. The 4 panel sub-figure on the right zooms into the boxed region at the type III burst site and shows the evolution of the weak jet as seen in AIA 94 \AA\ images. {The white star denote the centroid of the burst source shown in the left panel and the arrow points to the location of the RHESSI source. During the period between 06:13:01 -- 06:13:20 UT, when the first group of type III bursts were observed, no changes in the \B\ structure of the flaring region are seen.}}
    \label{mwa+aia}
    \end{figure}

\begin{figure}[!htb]
\centering
\includegraphics[width=0.85\textwidth,height=0.4\textwidth,clip=]{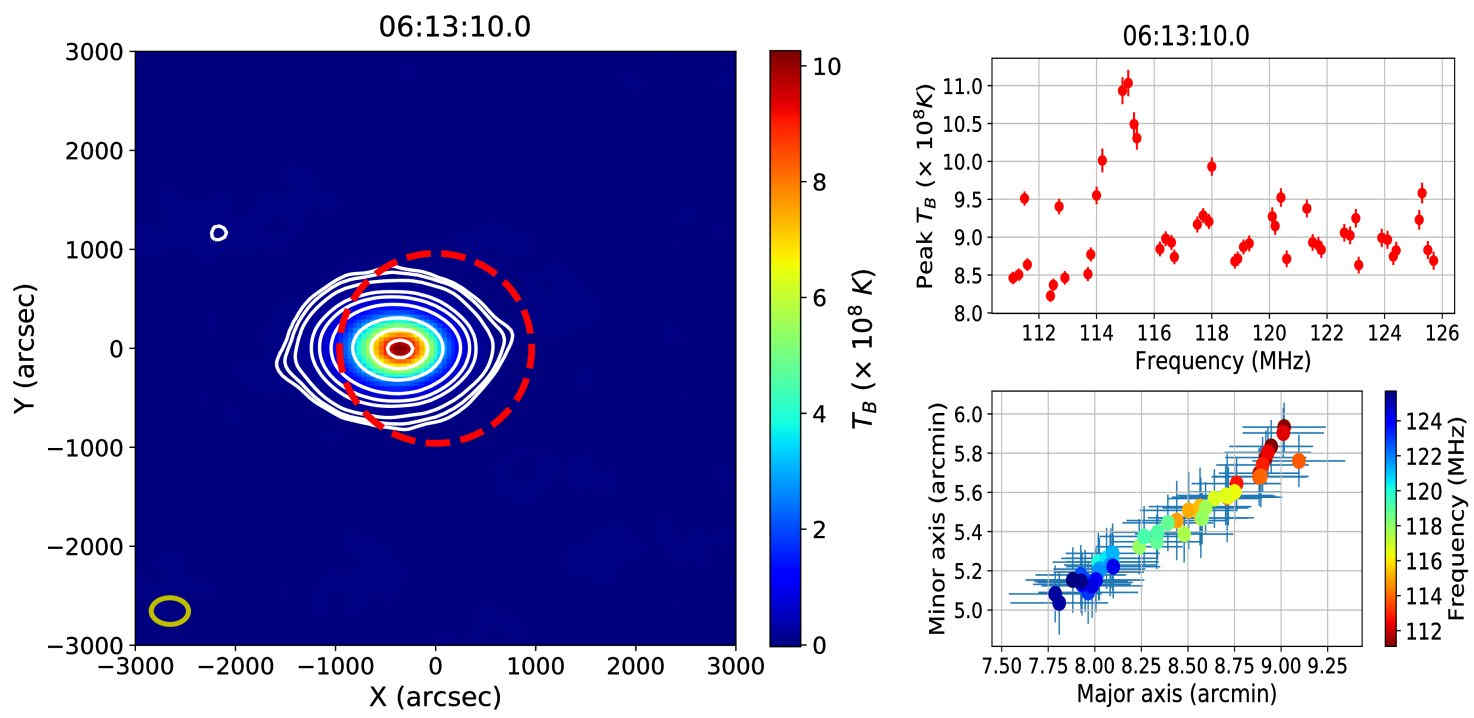}
\vspace{0.0001\textwidth}
\caption{{\it Left: }A sample \Tb\ map at 118 MHz and 06:13:10.0 UT. The contours are at 0.3\%, 0.5\%, 1\%, 3\%, 5\%, 10\%, 30\%, 60\% and 90\% levels of the peak \Tb. The dashed circle represents the optical solar disk and the ellipse in the lower left corner represents the FWHM of the synthesized beam ($8.2^{\prime}\times 6^{\prime}$). {\it Right: }The variation in the observed peak \Tb\ (top) and the major and minor axes (bottom) of the burst source as a function of frequency at 06:13:10.0 UT. These are the true source sizes after deconvolving the effect of the PSF. In the bottom panel, the spectral information is color-coded as shown in the colorbar. This snapshot is during the initial group of bursts.}
\label{gausfit}
\end{figure}

The entire 4 minute data set includes 6 groups of type III bursts. The first group of bursts started at 06:13:01 UT, and lasted for about 20 seconds. Following this, we observed intermittent groups of bursts separated by a few to a maximum of $\approx$ 20 s. The type III bursts were co-temporal and co-spatial with a weak active region jet which was visible in all observing bands of the Atmospheric Imaging Assembly (AIA; \citealp{Lemen2012}). Figure \ref{mwa+aia} shows an AIA 94 \AA\ full disk image of the Sun, close to the mid-time of the first group of bursts (06:13:01 -- 06:13:20 UT). The image is overlaid with MWA 118 MHz \Tb\ contours during the same time. The other panels in the figure zoom into the region of the jet marked by the box in the left panel, at different times starting from before the onset of the jet and start of the radio bursts to the end of the initial group of type III bursts. {The location of the centroid of the radio burst source at the mid-band frequency of 118 MHz during the mid-time of the first burst group is denoted by a white star in the panels. The arrow points to the location of the associated RHESSI source.}
It is evident from the AIA images that the magnetic loops retain their overall structure in spite of the weak jet. 
The other type III burst groups are also associated with similar weak jets from the same region. We find that the morphology of the burst source in radio maps was always modeled well by an elliptical Gaussian larger than the point spread function (PSF) corresponding to the observation. An example image is shown in Fig.\ref{gausfit} (left panel). We identified all images containing a compact and intense burst source (\Tb\ $> 10^8\ K$) among the entire set of images and obtained the best fit 2D Gaussian model for each of them. The model parameters included the Full Width at Half Maximum of the major and minor axes ($FWHM_{major(minor)}$), the position angle, the peak flux density and the location in sky coordinates. The effect of the PSF is deconvolved from the observed $FWHM_{major(minor)}$ to estimate the true source dimensions. Hereafter, $FWHM_{major(minor)}$ refers to the true source extents. In the right panel of Fig.\ref{gausfit}, we show the evolution of $FWHM_{major}$, $FWHM_{minor}$ and peak \Tb\ across frequency for a snapshot (06:13:10.0 UT). The goodness of the best fit model is evident from the very small error-bars in the estimates of these properties. The observed variations in these quantities are significant given the scale of respective uncertainties. For every burst image, using the values of Gaussian model parameters, we computed the integrated flux density, position angle relative to that of the PSF, standard deviations along its two axes ($\sigma_{major(minor)}$, computed as  $1/(2\sqrt{2\log 2}) \times FWHM_{major(minor)}$) and area defined as $\pi \sigma_{major}\sigma_{minor}$. The integrated flux densities were computed using the best fit peak flux density and area. 
All further analysis is based on the spectro-temporal evolution of source area, peak flux density, integrated flux density, relative position angle and sky location.
\subsection{The Evolution of best fit Gaussian source parameters}
\begin{figure}[!htb]
\centering
\includegraphics[width=0.95\textwidth,height=0.5\textwidth,clip=]{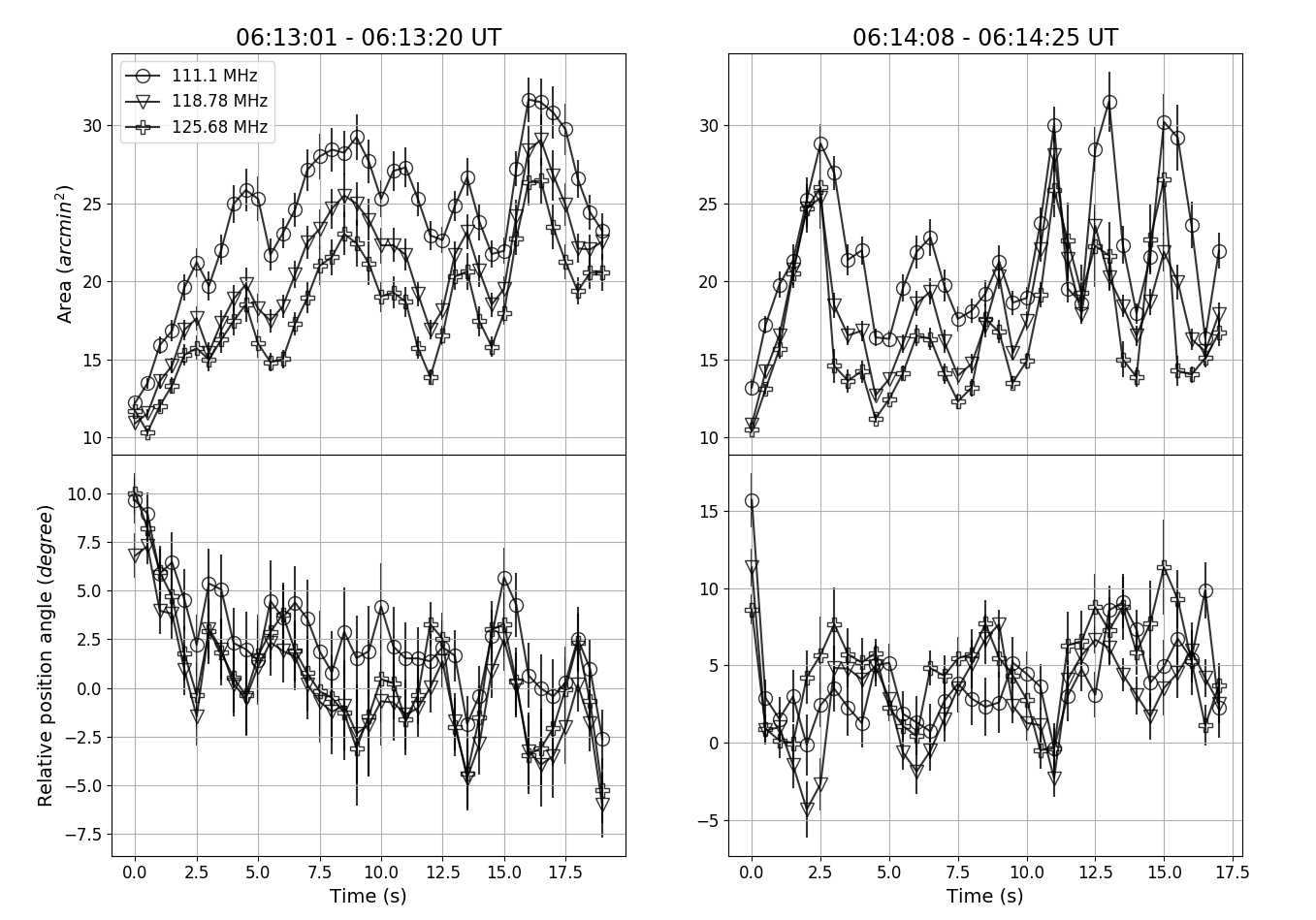}
\caption{{\it Top: }The temporal evolution of area of the source of burst emission for different frequencies during two different groups of bursts. {\it Bottom: } The evolution of relative position angle of the Gaussian source during the two burst groups. The frequencies are chosen to span the entire observation band. Note that the trends are qualitatively similar across frequency.}
\label{areaevol}
\end{figure}

In our discussion, we focus primarily on two groups of type III bursts. They span the time ranges from 06:13:01 -- 06:13:20 UT and 06:14:08 -- 06:14:25 UT. Hereafter, we refer to these groups as group 1 and group 2.
These are the top two long duration continuous groups of bursts and clearly show all the dynamical properties observed in the entire data set.  Figure \ref{areaevol} shows the temporal evolution of source area (top panels) and position angle relative to the PSF (bottom panels) for three different frequencies, during the two burst groups (left and right columns). 
It is evident from Fig.\ref{areaevol} that temporal evolution of source area and relative position angle show very significant variations which are qualitatively similar across the entire observing band. Seen most clearly in group 1, the burst source area shows a tendency to increase with time for about the first five seconds. Beyond this, the area tends to oscillate around a mean value in a quasi-periodic manner. This mean value lies between $20$ -- $25\  sq. arcmin$ and shows a dependence on frequency.

\begin{figure}[!htb]
\centering
\includegraphics[width=0.95\textwidth,height=0.28\textwidth,clip=]{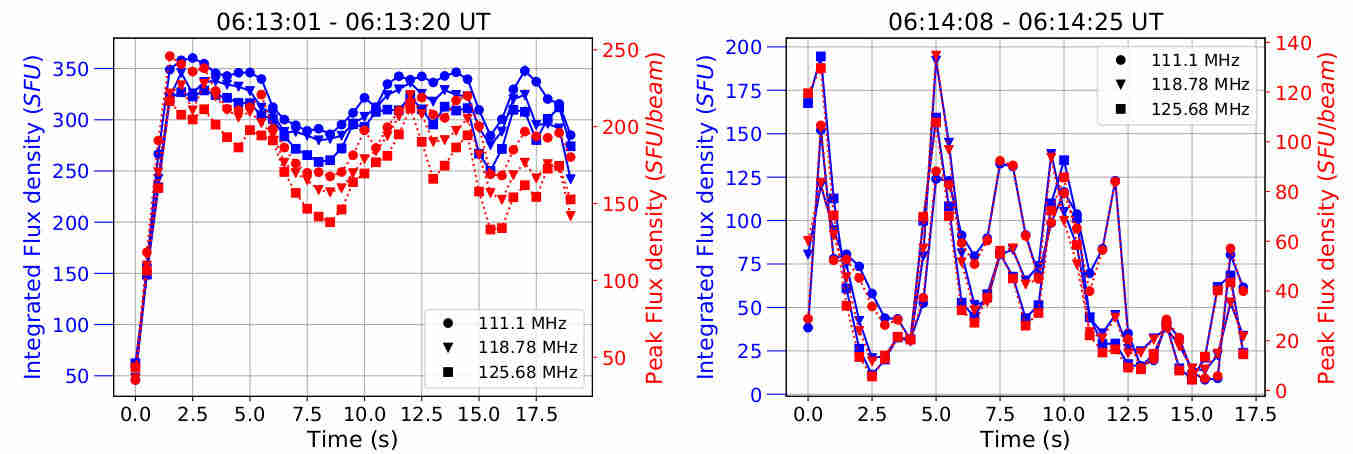}
\vspace{-0.01\textwidth}
\caption{{\it Left: } The evolution of integrated and peak source flux density for the two groups of bursts as a function of time, for different frequencies chosen to span the entire observation band. The trends found in time evolution are qualitatively similar across frequency for both the burst groups. There is a striking similarity in the temporal and spectral evolution of peak and integrated flux densities.}
\label{integPkevol}
\end{figure}

\begin{figure}[!htb]
\centering
\includegraphics[width=0.8\textwidth,height=0.45\textwidth,clip=]{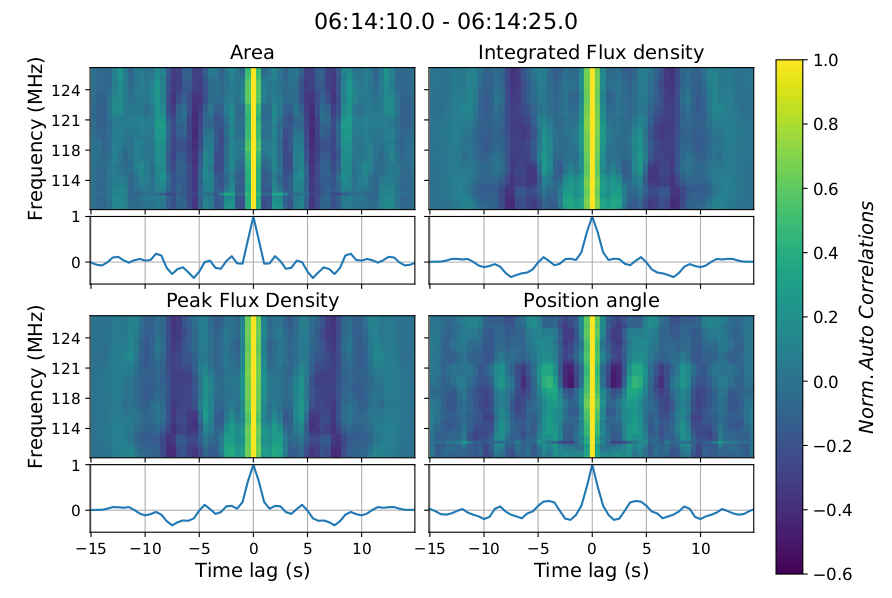}
\vspace{-0.001\textwidth}
\caption{Normalised auto-correlations of area, integrated flux density, peak flux density and relative position angle during the bursts from 06:14:10 -- 06:14:25 UT. The time window for the analysis was so chosen that the burst properties had already evolved to a stable mean value. Note that the evolution of properties are similar across frequency. The line plot at the bottom of each panel shows the normalized auto-correlation function averaged across frequency. The second-scale pulsations are evident in these plots.} \label{NAC}
\end{figure}

Figure \ref{integPkevol} shows the time evolution of integrated and peak flux density for the source of burst emission for the same two burst groups, for the same frequencies as used in Fig.\ref{areaevol}. Once again, very similar trends are observed across the band. The integrated flux density and peak flux density trends closely follow each other. Integrated and peak flux densities are observed to change by factors of many over a matter of seconds.
 Like the area evolution, the integrated flux density shows a dramatic linear increase at the start of the first group of bursts (left panel) and show strong QPOs mainly during the second group of bursts (right panel). 
 To study the periodicity in the evolution of various properties, the normalised auto-correlations were computed for all the properties (Figure \ref{NAC}).
The vertical bands in the figure imply that similar behavior is seen at all frequencies. It is evident that the oscillations in area have a timescale of $\approx 2 s$. 
However, the relative position angle has a slower quasi-periodicity of $\approx 3s$.   

\begin{figure}[!htb]
\centering
\vspace{-0.001\textwidth}
\includegraphics[width=0.9\textwidth,height=0.3\textwidth,clip=]{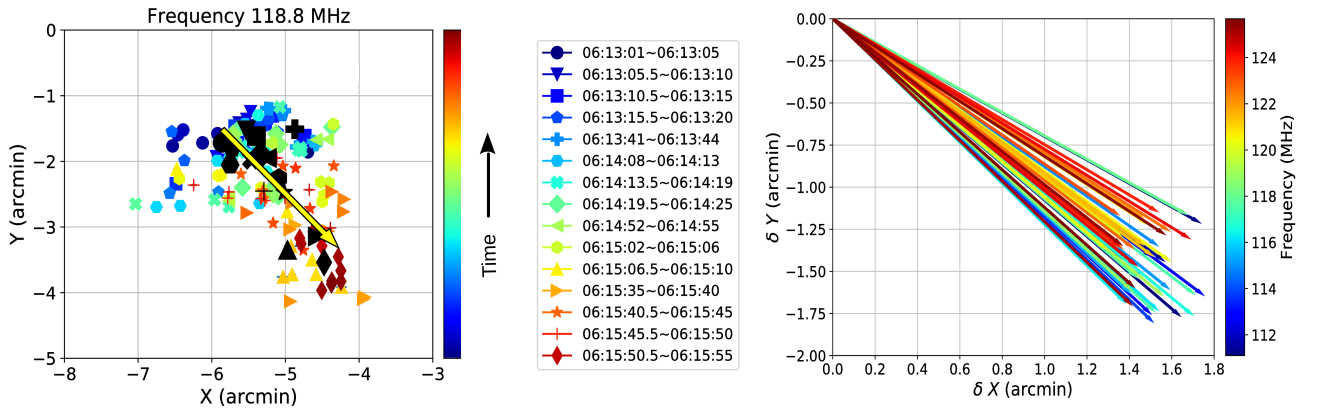}
\caption{{\it Left: }The source locations of type III bursts observed at 118.8 MHz during the entire observation period. Each marker indicates the time period corresponding to the burst and the marker color has a gradient showing the variation in event occurrence time. Blue labels correspond to earlier bursts and red the ones towards the end of observation. The bursts are grouped based on occurrence time into intervals of 5 s duration. Black markers correspond to the mean location of the burst sources observed during the respective intervals. The yellow arrow is the {\it mean source drift vector}, obtained as a linear fit to the black points. {\it Right: } {The mean drift vectors obtained for all the imaging frequencies, with their tails aligned to start from the same point. They tend to lie within a narrow cone and don't show any trend in frequency.}}
\label{locdrift}
\end{figure}

Temporal variation in the location of burst sources have been reported by earlier authors (\citealp{Duncan1985}; \citealp{kontar2017}). Figure \ref{locdrift} (left panel) shows the observed locations of the burst sources at 118.8 MHz, close to the centre of the observation band. The plot contains information of all the bursts observed during our observation period. The bursts were sorted according to the increasing order of their occurrence time and grouped in 5 second intervals leading to 15 groups. The plot legend gives the details of the time period of each group and the allocated marker. Black markers in the plot represent the {\it mean location} of the burst sources observed in each 5 s interval. The yellow arrow is the {\it mean source displacement vector} (hereafter `drift vector'), obtained by a linear fit to the mean locations. 
{The drift vectors for different spectral channels lie at nearby but different locations in the image plane, though their lengths and orientations are very similar. 
The right panel shows the drift vectors for all of the imaging frequencies aligned such that their tails coincide.}
We find that the location of the burst sources show a systematic drift in the sky plane which is quite similar across the observation band. All the drift vectors lie within a narrow cone towards a particular direction. They don't show any significant trend in their magnitude or direction with frequency. The mean magnitude and direction of the drift were found to be ${2.09 \pm 0.693}^{\prime} $ along $\approx {-44.7}^0$ spread within a cone of opening angle $18^{0}$. The observed magnitude of the drift corresponds to $\approx 90.8 \pm 30.1\ Mm$ in projected distance.

\section{Interpretation}\label{discussion}
From the discussion in the previous section, it is evident that the properties of the burst source evolve systematically and significantly across time. The temporal evolution is qualitatively similar across the narrow band of observations. {As discussed in Section \ref{intro}, since we observe only a single streak of coherent emission in the Culgoora dynamic spectrum (Fig.\ref{culg+mwa}), we assume that this is the harmonic emission component.} Adopting the density model by \cite{zucca2014} (hereafter, Z model), the width of the coronal region explored by the 15 MHz band is $\approx 21 Mm$, less than $10\%$ of the hydrodynamic coronal scale height. This explains the qualitative similarity in the temporal evolution of source properties across our observing band. Note that correlation scales of local density fluctuations can be as small as 0.1 Mm, effectively introducing fluctuations from channel to channel in our 160 kHz resolution images (\cite{Arzner1999}; \cite{sasi2017}). This is a plausible explanation for the small differences observed. 
Since, the temporal evolution has been shown to be qualitatively similar across frequencies, in the following discussion we focus on a single frequency of 111.1 MHz. For a Gaussian source model, the area, integrated flux density and peak flux density are not all independent quantities and between them have only two degrees of freedom. As has been shown (Fig.\ref{integPkevol}) the trends in the peak and integrated flux densities follow each other closely. In the following discussion, we work with integrated flux density, which is a measure of total energy radiated and remains unaffected by coronal radio-wave scattering. 
\subsection{Co-evolution of integrated flux density, burst source area and relative position angle}
\begin{figure}[!htb]
\centering
\includegraphics[height=0.3\textwidth,width=\textwidth,clip=]{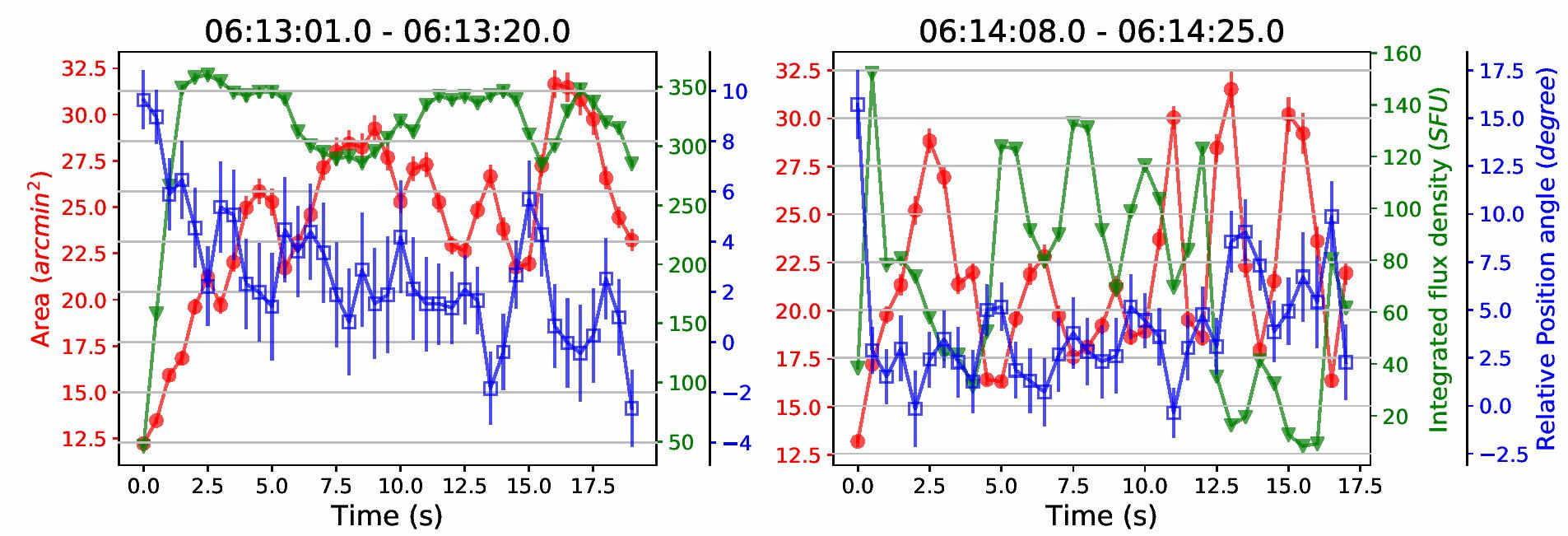}
\caption{The two panels show the evolution of integrated flux density ({\it triangle}), area ({\it dot}) and relative position angle ({\it square}) of the Gaussian sources during two groups of type III bursts at 111.1 MHz.}
\label{Areaintegevol}
\end{figure}

\begin{figure}[!htb]
\centering
\includegraphics[width=0.85\textwidth,height=0.6\textwidth,clip=]{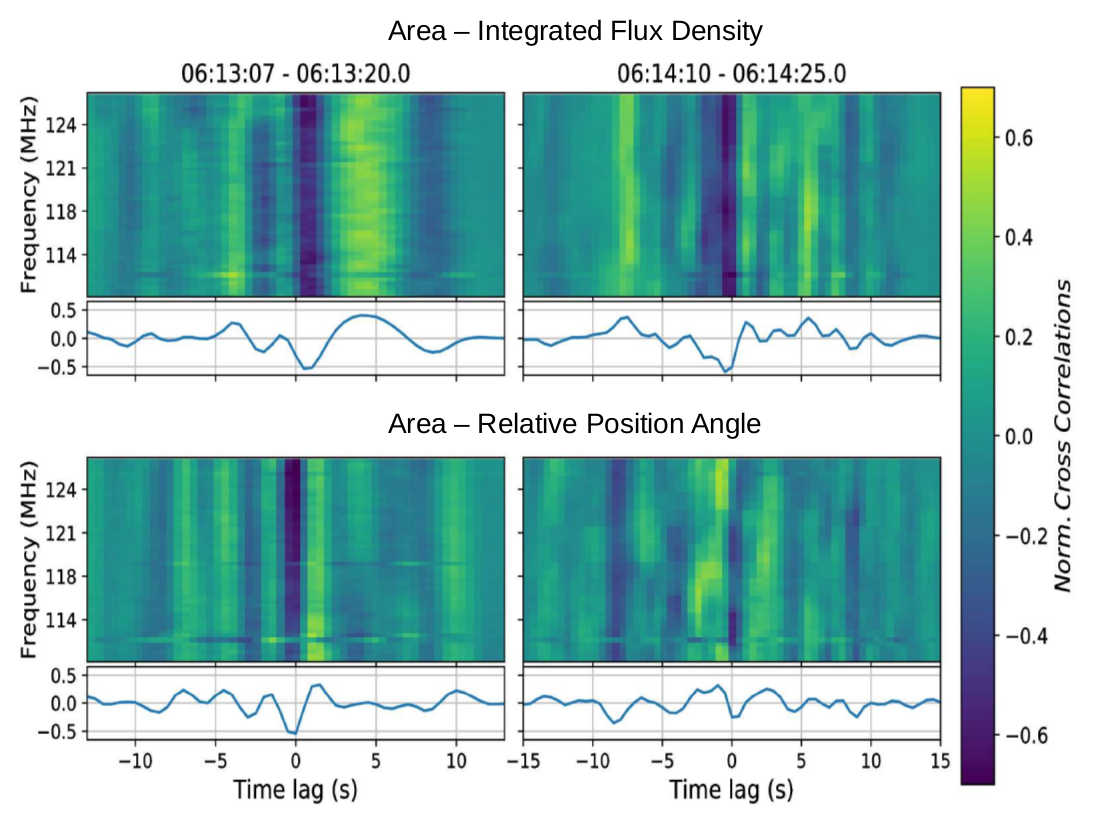}
\vspace{0.0001\textwidth}
\caption{{\it Top: }Normalised cross-correlation matrices for both the burst groups. The analysis excludes the initial rise phase for both the burst groups and the corresponding time periods are provided in the respective titles. Each column is the normalised cross-correlation between integrated flux density and area time series for a particular spectral slice. The x axis shows the time lag. A strong anti-correlation is evident close to zero time lag. The vertical stripes arise due to the oscillatory nature in the time series data. The plot in the smaller panel below each subplot shows the mean normalised cross correlation across frequency. The peak anti-correlation is around -0.6. 
{{\it Bottom: } Normalised cross correlations for area and relative position angle in the same format.
While the first group of bursts shows a significant anti-correlation at zero time lag, the second group shows multiple peaks and valleys of lower significance and similar strengths at various lags.
}}
\label{Areainteg_cc}
\end{figure}

Figure \ref{Areaintegevol} shows the co-evolution of integrated flux density, area of the source of burst emission and position angle during the two chosen groups of bursts. The integrated flux density can be regarded as a proxy for the particle energy density in the electron beam that generated the two stream instability and excited the coherent Langmuir waves (\citealp{melrose2017}), a part of which eventually was converted into coherent electromagnetic waves. 
The left panel of Fig.\ref{Areaintegevol} shows the co-evolution of integrated or the net coherent flux density, area and relative position angle of the burst source during the first group of type III bursts. The rise in the net coherent flux density is sudden while the area of the source shows a more gradual rise and tends to level off around $\approx 25\ sq. arcmin$. The right panel shows the co-evolution of the same parameters during the second group of bursts. There also, the area  levels off to a value in the vicinity of $ 25\ sq. arcmin$. After reaching this mean value, the area shows QPOs antiphased with the oscillations in the net coherent flux density. This behaviour is seen more prominently in the right panel of Fig.\ref{Areaintegevol}. Figure \ref{Areainteg_cc} shows the normalised cross-correlation of time evolution of integrated flux density and relative position angle with that of the area of the burst source during the QPO phase, across the observation band.
{A strong anti-correlation is observed between area and coherent flux density for both groups of bursts. The near periodic oscillations away from zero time lag in the normalised cross correlations arise due to the quasi-periodic variations in these properties. The anti-correlation peaks at a time lag of $\approx 0.5 s$ for the initial burst group and $\approx -0.5 s$ for the second group. 
In view of the $0.5 s$ time resolution of these data, this observed lag is insignificant. For the correlation between area and relative position angle, the two groups of bursts show different behaviors. While the first group of bursts show an anti-correlation of significance similar to those seen between area and integrated flux density, the second one shows multiple weaker peaks of either sign and at multiple lags. Extending similar analysis to rest of the burst groups, we find that all of them show anti-correlation of similar magnitude between area and coherent flux density.
On the other hand, different groups of bursts show different cross-correlation trends between the area and the relative position angle, with the normalized cross-correlations spanning the range of $\pm 0.5$ close to zero lag. Hence, we cannot make a strong conclusion about the correlation between these two properties.
The similarity in the normalised cross-correlation values across frequency is obvious in Fig.\ref{Areainteg_cc} and constitute evidence for qualitative similarity in temporal variation of burst properties across frequency. We find this trend for the rest of the burst groups as well.}
\subsection{Area growth}
\begin{figure}[!htb]
\centering
\includegraphics[width=0.84\textwidth,height=0.32\textwidth,clip=]{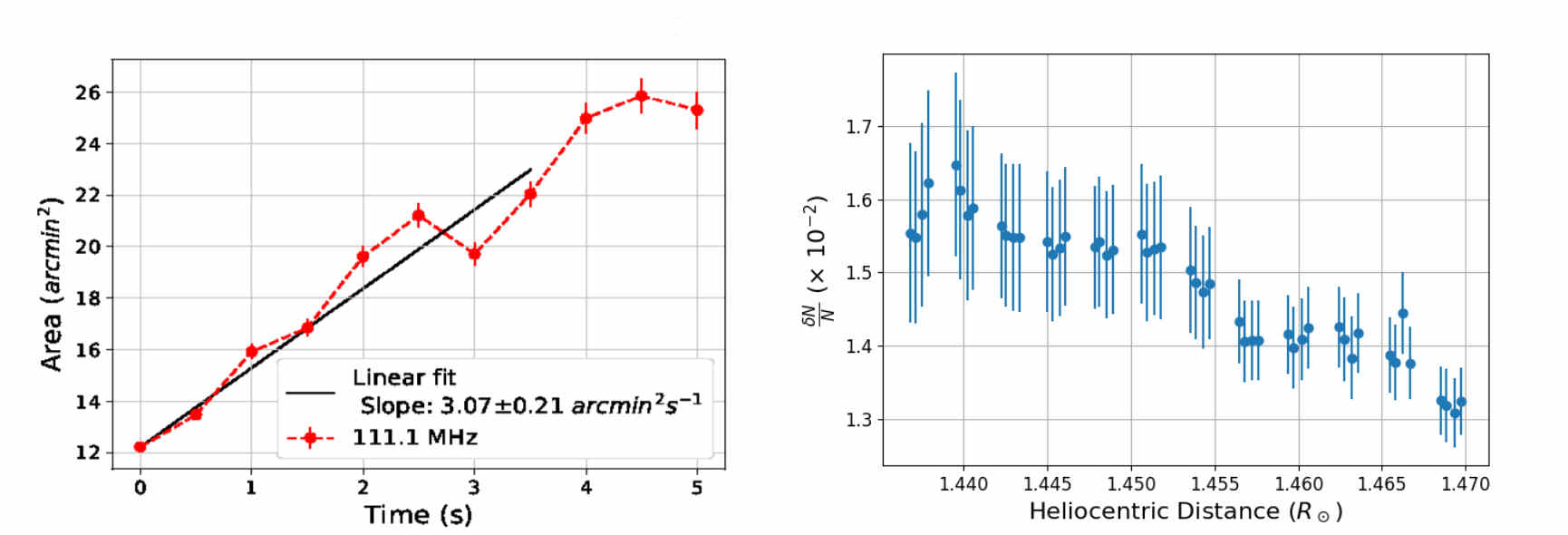}
\caption{{\it Left:  }Diffusive growth period of source area at 111.1 MHz for the first group of bursts. Linear fit is done to estimate the diffusion rate. {\it Right: }\dn\ as a function of heliocentric distance calculated using the area diffusion coefficients derived across frequency. These values are nearly four times larger than the theoretical estimate for ${\delta N/N}_{sat}$. Refer the text for details.}
\label{DNest}
\end{figure}

{A linear fit to the steadily growing phase of the area of the burst source reveals a growth rate of $5800\ Mm^2/s$, i.e a radial expansion of $\approx 43\ Mm/s$ (Fig.\ref{DNest} left panel). This is nearly two orders of magnitude larger than the local \alf speed (typically $0.5\ Mm/s$) and hence cannot be interpreted as an increase in physical area due to any magnetic field or plasma dynamics. A similar observation was reported by \citet{kontar2017} at lower observation frequencies.} This phenomenon is predicted by the theory of ray propagation through a stochastic plasma (\citealp{steinberg1971}; \citealp{robinson_scat1983}; \citealp{cairns1998}; \citealp{arznerPHD} ). Here, we use the formalism developed by \citet{Arzner1999} (AM hereafter) based on a ray optics approach to model propagation effects observed at the metre-waves. They derived an expression linking the phase space $(\vec{x},\vec{k})$ diffusion coefficient ($\eta^*$) of the radiation to an observable -- the growth rate of area of the source, $D_s$. AM predict a linear growth of burst source area followed by a saturation. The saturation area is a function of the strength of density fluctuations, \dn, where $N$ is the mean local electron density and $\delta N$ is the local departure from its mean value. \dn\ is expected to remain unchanged over timescales of order a few minutes. The saturation area increases with \dn. However, there exists a saturation strength of density fluctuations, ${\delta N/N}_{sat}$, beyond which saturation area asymptotes to a constant. The regime of turbulence, where \dn\ $>\ {\delta N/N}_{sat}$ is termed as the strong turbulence regime, where even a large increase in \dn\ cannot produce significant variation in the size of the scattered source. In this regime, the size of the scattered source asymptotes to $\sigma_{max}$. $\sigma_{max}$ is independent of \dn, but depends on the thickness of the scattering medium, $L$. Another consequence of scattering is that the rays propagating through the strongly turbulent region get delayed. The average time taken by the photons to exit the scattering medium, $<t>$, can also be estimated in terms of properties of the medium. To facilitate the following discussion, we provide the relevant equations below from AM.
 \begin{eqnarray}
     \sigma_{max} &=& \frac{L}{\sqrt3 D}  \label{sigmax} \\
     {\delta N/N}_{sat}  &=& 1.5 \frac{{\overline{n}}^2}{1-{\overline{n}}^2}\left( \frac{l_i}{L} \right)^2  \label{dnsat} \\
     <t> &=& \eta^*\left(\frac{L}{c\overline{n}}\right)^2   \label{peaktime}\\  
    \eta^* &=& \frac{\pi}{8} c<\kappa> \frac{\omega_p^4}{\omega^4}\left(\frac{\delta N}{N}\right)^2 \label{eta}\\ 
     D_s &=& \frac{\pi(c \overline{n})^2}{3\eta^*D^2}  \label{diff}
 \end{eqnarray}    

where, $<\kappa>$ is the mean scale of fluctuations averaged over the turbulence spectrum, $D$ is the Earth - Sun distance, $\bar{n}$ is the mean refractive index within the scattering region of thickness $L$ and $c$ denotes the speed of light. If one assumes Kolmogorov turbulence it can be shown that $<\kappa> \approx 1/l_i$, where $l_i$ is inner scale length of turbulence. The best value for $l_i$ was shown to be three times the ion inertial length by \citet{coles1989}, using various observations that probed $\approx 2 - 22\ R_\odot$ region. The original equations derived by AM were for the evolution of the variance of the 2D Gaussian source, $\sigma^2$, in spatial coordinates. We compute source area as $\pi \sigma_{major} \sigma_{minor}$ since our burst sources are elliptical Gaussian shaped and the co-ordinates are angular sky-coordinates. Taking these aspects into consideration, the equation for $D_s$ has been modified. 

Linear fits to the rising part of the area evolution curves for the first group of bursts were performed and $D_s$ was estimated across our observation bandwidth. Figure \ref{DNest} (left panel) shows such a linear fit for the area evolution curve at 111.1 MHz. We note that the integrated flux density takes $\approx 2s$ to level off or attain its `saturation value' (Fig.\ref{integPkevol}). This is an estimate of the mean intensity saturation time, $<t>$ and is found to be constant across the observing band. To calculate \dn\ we need to get an estimate of $L$ and \n. However, \n\ and $L$ are not independent quantities. If we have an estimate for $L$, \n\ within the scattering medium can be obtained, because \n\ is a function of $L$ and the coronal density function (Z model in our case). Knowing $D_s$ and $<t>$, we can combine Eqs. \ref{peaktime} and \ref{diff} to eliminate $\eta^*/$\n$^2$ and compute $L$. $L$ was estimated for every frequency slice, using which \n\ was computed. Next, using Eqns. \ref{eta} and \ref{diff} we estimated \dn\ as a function of frequency. Figure \ref{DNest} shows the \dn\ estimates for all observation frequencies. The frequencies are converted to coronal heights using the Z model. The values obtained for ${\delta N/N}_{sat}$ using Eqn. \ref{dnsat} are around $4 \times 10^{-3}$. Though they involve the choice of a particular density model, this approach has allowed us to estimate \dn\ at arguably the lowest coronal heights. Our \dn\ estimates are twice that obtained by \citet{mugundhan2017} at a height range of $\approx 1.6$ -- $2.2\ R_\odot$.    
It is worth mentioning that the values of $L$ obtained for different frequencies are typically $\approx .1\ R_\odot$ which is larger than the range of heights spanned in Fig.\ref{DNest}. This is the likely reason for not finding any strong variation in \dn. The estimated \dn\ is much higher than ${\delta N/N}_{sat}$, implying a strong scattering regime. Hence, as predicted by AM, ray propagation becomes diffusion dominated which results in the observed linear growth in burst area. The predicted saturation in the area is also observed following the linear growth phase, which yields an estimate of the asymptotic saturation area $\sigma_{max}$.

{Saturation area was computed using Eq. \ref{sigmax}. A $L = .1\ R_\odot$ leads to a saturation area, $A_{sat,theory}$, of  $5\ sq.arcmin$, which translates to an FWHM of $2.9^{\prime}$. This theoretical estimate is in good agreement with the Monte-Carlo simulation result by AM and \citet{steinberg1971}.
The latter carried out a detailed study estimating the size of a scattered point source as a function of  \omgp $/\omega$, longitude,  coronal height of the source and $\delta N/N$.
For a source located at a height of 1.2$R_{\odot}$, corresponding to an  \omgp $/\omega \approx .68$, and a longitude of 45$^{\circ}$ surrounded by a turbulent medium with $\delta N/N$ of $2\%$, they computed a FWHM of 3$^{\prime}$ at 169 MHz. The various parameters of this simulation are close to the situation encountered here.
Figure \ref{Areaintegevol}, however, shows that the area saturates roughly around $A_{sat,obs} = 25\ sq.arcmin$, about 5 times larger than the expected value of $A_{sat,theory}$.
 The observed deconvolved source area of $25\ sq.arcmin$ corresponds to an FWHM of $5.9^\prime$.
 The observed sizes of sources of burst emission are consistent with earlier reported values \citep[e.g.][and many others]{sheridan1972, dulk1980}. {\citet{RA} reports a typical value of $5^{\prime}$ for the deconvolved size of type III burst sources at 169 MHz reviewing  previous observations.} \citet{Hilaire} studied nearly 10,000 type III bursts observed with the NRH over a decade and reported a mean source FWHM of $5.3^\prime\pm1.8^\prime$ at 150.9 MHz. 
Given the coherent emission mechanism widely accepted to be at play, the large observed angular sizes of type III sources have been a puzzle.
This has lead to the suggestions that the sources of type III emissions must themselves comprise multiple sites of emission \citep[e.g.][]{mercier75_multiBstrands_typeIIIsource, raoult1980, Pick1980, Duncan1985}.
The large intrinsic sizes of sources of burst emission from this work provide further evidence for this possibility.
} 

To proceed, we assume that the response of the scattering medium to the actual burst source is described by the convolution of the intrinsic source morphology and the theoretical saturation area, $A_{sat,theory}$, representing the response of the medium to a delta function source. Having estimated $A_{sat,theory}$, we can then derive the true size of the burst source. This suggests a source of elliptical Gaussian geometry, which is reasonable given that the cross section of a flux tube will in general be an ellipse and that the current density of the accelerated electron beam is expected to gradually fall off towards its edges (\citealp{roussev2001}; \citealp{Ash}).  We calculated the size of the true Gaussian source, \sigsrc\ under this scenario, using the expression, $ \sigma_{source} = \sqrt{ \left[ (A_{sat,obs} - A_{sat,theory})/ \pi \right]} $. The source size was found to be $2.5^{\prime}$, or approximately $0.15\ R_\odot$. \citet{Asch_observersview2003} gives an empirical estimate for the coronal loop width, $a$, as $\approx 0.2h$, where $h$ is the height of the loop. In our case, $h \approx .45\ R_\odot$, which gives, $a\approx 0.09\ R_\odot$. Given that the open flux tubes are more divergent magnetic structures than closed loops, our estimate of $\sigma_{source}$ seems reasonable.

{These observations have been established to be in the strong scattering regime. 
The discussion till now has focused on the impact of scatter on the observed size of a source.
The multi-path scattering also leads to temporal broadening of any time variable source (\citealp[e.g.][]{steinberg1971,riddle74_pulseBroaden_scatt}).
The impact of temporal broadening can be modeled by a convolution with the impulse response function corresponding to the propagation medium.
We note that while this will tend to reduce the observed amplitude of the QPOs, it will leave the period unchanged.
}
\subsection{Quasi-Periodic Pulsations in Area, Integrated flux density and relative position angle}
As presented earlier, we find QPOs in the area of the burst source and integrated flux density or the net coherent flux density and the orientation of the elliptical source (Fig.\ref{Areaintegevol}). There is a notable anti-correlation between the first two properties (Fig.\ref{Areainteg_cc}) as well. {We explore the following possibilities that can lead to the observed anti-correlation.}
\subsubsection{A variable intermittent compact source co-located with an extended persistent source: }\label{varsrc}
{A possible scenario which can give rise to the observed anti-phased QPOs in source area and integrated flux density is as follows. Assume a persistent extended background source at the location of the bursts. The type III sources appear as compact intermittent variable flux density sources over this broad background source. When fitted with a single 2D Gaussian, such a model will show a larger area for the source when the burst source is weak or absent, and a smaller area when the emission is dominated by the compact intermittent burst source. This will naturally give rise to the observed anti-correlation between the area and the integrated flux density of the source. We have examined this possibility in detail.
Our analysis shows that the hypothesized two {\em independent} features must be spatially coincident, appear/disappear at the same time, and undergo very large absolute variations in strength in a similar pattern while maintaining comparable {\em relative strengths}.  In other words their gross properties are extremely highly correlated with each other, and they only differ in terms of short-term relative flux densities. 
This is strong evidence that the two hypothesized phases (strong/compact and weak/extended) are simply different states of a single phenomenon, and not two different phenomena superimposed by chance.
The details of this analysis are presented in Appendix \ref{app2}.
}
\subsubsection{Scattering: }
If these variations in the area of the source arise due to scattering effects (while the intrinsic source size remains unchanged), it would require corresponding variations in \dn. However, our \dn\ estimates already place us in a strong scattering regime, where the observed area of the source tends to  saturate and is not sensitive to changes in \dn. \dn\ is also not expected to vary at a few second scales. Hence these variations are not likely to be due to scattering effects. 
\subsubsection{Local MHD oscillations: }
A potentially convenient way to produce these variations in area, accompanied by the anti-phased variations in intensity is as follows. The theory of generation of coherent emission at burst sites says that the energy density in the coherent Langmuir waves is given by, $W_L = \Gamma n_b m_e {v_b}^2$, where $\Gamma$ is the efficiency factor, $n_b$ is the electron beam number density, $m_e$, the electron mass and $v_b$, the beam velocity (\citealp{melrose2017}). The easiest way to vary the coherent energy density is by changing $n_b$, which is easily done by modulating the `width' of the magnetic flux tube guiding the electron beams. Though the bulk velocity, $v_b$ of the beam electrons are very high, they remain bound to the magnetic field lines within the open flux tube owing to low plasma $\beta$ in the corona, and the fact that most of the bulk speed of these electrons is directed along the \B\ direction. 
It is tempting to argue that fast sausage mode MHD oscillations in the open flux tube can generate such modulations in the beam number density leading to such QPOs in the integrated or the net coherent flux density. In fact, many of the previous reports of second-scale QPOs in radio burst intensity (\citealp{tapping1978}; \citealp{Chernov1998}; \citealp{kattenberg1983_muwaveTypIII_QPPmodel}) were interpreted based on MHD waves or related phenomena (\citealp{rosenberg1970}; \citealp{roberts_fastQPP1984}; \citealp{Ash2004_SausMode_NthHarmonic_QPO}). However, the rapid large scale motion implied by the second-scale oscillations of a coronal flux tube of radius $0.15\ R_\odot\ \approx 100\ Mm$ imply speeds of $\approx 100\ Mm/s$. However, the typical local \alf speed in these regions is expected to be $\approx 0.5\ Mm/s$, using magnetic field estimate from \cite{dulkml78} and the density estimate from the observation frequency assuming harmonic emission. Combined with the observation that the QPOs are coherent across the $21\ Mm$ height range (corresponding to 15 MHz band) within which typical information travel time would be 40 s assuming the typical \alf speed, we rule out the models based on local MHD processes for the second-scale QPOs.
\begin{figure}[!htb]
\centering
\includegraphics[width=0.9\textwidth,height=0.35\textwidth,clip=]{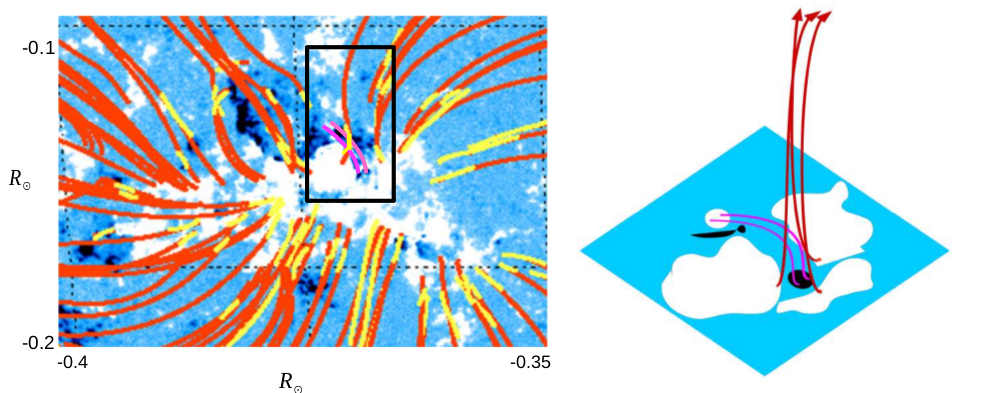}
\vspace{0.0001\textwidth}
\caption{{{\it Left:} VCA-NLFFF extrapolation of the region around the AIA jet. Red lines show the extrapolated field lines. The yellow lines are the loop strands identified in AIA images which coincide with the extrapolated field lines. The magenta lines inside the black box identify the loops seen in AIA maps which were bright during the jet event. {\it Right: } A schematic representation of the event. Black and white islands denote opposite magnetic polarities. Red lines are open field lines, while the magenta lines show the closed loop.}}
\label{Bscheme}
\end{figure}  
\subsubsection{Modulation of particle acceleration/injection}
A way to make the electron beam flux density oscillate, without running afoul of the \alf\ speed restriction is to modulate it at the site where the electrons are accelerated/injected in the flux tube. This would require the presence of a quasi-periodic regulation mechanism operating at the accelaration/injection site. 
{Figure \ref{Bscheme} shows the magnetic field structure at the region around the jet obtained using a Non-Linear Force Free Field (NLFFF) Extrapolation code by \cite{ash16_vcanlfff}, based on Vertical-Current Approximation (VCA). 
The yellow lines show the field lines which could be identified in the AIA images, and the extrapolated field lines are shown in red.
The background image shows the line of sight magnetogram map from Helioseismic and Magnetic Imager (HMI; \citealp{Scherrer12}).
The site of the jet is marked by the black box. The magenta field lines within the box show the loops that underwent reconnection and produced the jet in Fig.\ref{mwa+aia}. Though they are seen clearly in the AIA images, these small closed loops were not reproduced in the extrapolation. They were hence marked manually by comparing AIA images with magnetogram data. 
A schematic diagram of the region of interest, based on this model is shown in the right panel of the figure.
A comparison with Fig.\ref{mwa+aia} suggests that it is one or both of the magenta closed loops which intermittently reconnected with the neighboring open field lines to produce the jet seen in AIA images.
This is very similar to a standard jet event \citep{moore2010}.
Assuming that the intermittent reconnection, giving rise to the observed groups of type III bursts, happened at the loop top region, we can estimate the MHD oscillation period.
A circular geometry assumption for the closed loop gives the height of the loop top to be $3.5\ Mm$, which corresponds to an electron  density of $2\times 10^9 cc^{-1}$ using the Z model.
The NLFFF extrapolation yielded a magnetic field around $261\ G$ near that region. 
These estimates lead to a typical \alf speed of $12.75\ Mm/s$.
Assuming that the reconnection events happen on both the loop tops, the maximum thickness of the particle injection sites would be twice the loop width. Using the empirical loop width-to-length rule mentioned earlier, the loop top widths are around $2.8\ Mm$ . 
Hence QPOs with periods as fast as $0.5\ s$ do not pose a challenge for MHD processes. 
The idea of modulating the non-linear reconnection and particle acceleration process leading to quasi periodic oscillations in the jet has been proposed earlier \citep{Asch1994_coherent_or_randphase_Acclbeam} and form the third class of models described in Section \ref{intro}.

However, these models focused on explaining the QPOs in intensity alone as the simultaneous oscillations in radio source size and orientation had not been discovered. 
In the light of our observations, we put forth the following qualitative picture.
Consider a reconnection layer between a coronal loop and open field lines.
{Closed loop structures are known to harbour various MHD wave modes. 
As the reconnection layer is compressed externally by a MHD wave mode, the magnetic field gradient across the separatrix increases leading to a faster reconnection rate and a thinner reconnection layer \citep{rutherfor73_tearmodewidth}. 
These lead to an increase in the injected electron flux density. The compressed magnetic field structure about the separatrix and the consequent reduced width of the reconnection layer simultaneously leads to a decrease in the beam solid angle. The reverse happens as the layer is relaxed by the wave mode. This leads to the anti-correlation between the solid angle and flux density of the electron beam. In radio images this manifests as anti-phased QPOs between area of the source of burst emission and its integrated flux density.}
It is possible that the QPOs in source orientation are associated with a different MHD mode, since it does not show any evidence for a persistent sense of correlation with other properties. 
This different mode could perhaps be a torsional mode. 
A more detailed analysis or modelling is beyond the scope of this work.
} 
\subsection{Source drift}
Though superposed with a seemingly random scattering based component, bursts sources showed a clear systematic motion in the sky plane with time. Figure \ref{locdrift} shows the observed systematic drift in the burst source with time. The typical time taken for this drift is $\approx 3\ minutes$, the duration from the start of the first group of bursts to the end of the last group. We estimate the average drift speed of the burst sources to be $\approx 0.5 \pm 0.17\ Mm/s$. Using the model by \citet{dulkml78}, we calculated the magnitude of \B\ at our coronal heights, and the number density corresponding to the mid-band frequency of 118 MHz to estimate a typical \alf speed, $V_{Ae}$. $V_{Ae} \approx 0.55\ Mm/s$, which is close to the observed source drift speed. This suggests that the overall source drift could be due to the motion of the open flux tube within which the bursts emission arises. {The large scatter around the mean source locations marked by black markers in Fig.\ref{locdrift} is likely a result of scattering.} A detailed study of apparent source positions due to scattering was done by \citet{robinson_scat1983}, using Monte-Carlo simulations. The authors showed that the apparent positions can be shifted by order of a few arc-minutes for the harmonic emission component from the burst sites, which are located within open flux tubes. The observed scatter in our burst source locations is consistent with this. 

\section{Conclusion}\label{conclusion}
In this work, we have studied in detail the radio emission from groups of weak type III bursts associated with faint active-region jets using snapshot spectroscopic imaging. Using these observations from the MWA and a framework  to describe ray propagation through the turbulent coronal medium, we demonstrate a method to disentangle the effects of radio wave propagation from the intrinsic dynamics at the burst site. We report the discovery of second-scale QPOs in the source size and sky orientation simultaneous with pulsations in intensity. Using a reasonable coronal density model and assuming that the observation frequency is the first harmonic of local \omgp, the height of these sources is estimated to be around $1.45\ R_\odot$. The observed QPOs correspond to $\approx25\%$ variations in $\approx0.15\ R_\odot$ wide sources coherently across a height range of $0.03\ R_\odot$. Sustaining these oscillations within a MHD framework would require a local \alf speed two orders of magnitude larger than typical at these coronal heights. We, hence, rule out the role of local MHD waves or related phenomena in second-scale  QPOs associated with type III bursts. Our observations not only provide direct evidence for a quasi-periodic regulation of the particle injection, but also give the first observational indications of modulations in the solid angle and particle flux density of the electron beams that trigger the radio bursts. Additionally, we estimate the density fluctuations in the quiet corona at a heights of $\approx 1.45\ R_\odot$, which is the first of such estimates in the mid-Corona using direct observations of source scattering. This approach provides a tool which can potentially routinely be used to characterise and quantify mid-Coronal plasma turbulence using the frequent weak radio burst events.

\section{Acknowledgements}
This scientific work makes use of the Murchison Radio-astronomy Observatory, operated by CSIRO. We acknowledge the Wajarri Yamatji people as the traditional owners of the Observatory site. Support for the operation of the MWA is provided by the Australian Government (NCRIS), under a contract to Curtin University administered by Astronomy Australia Limited. We acknowledge the Pawsey Supercomputing Centre which is supported by the Western Australian and Australian Governments. The AIA data are courtesy of SDO (NASA) and the AIA consortium. The SDO/HMI data were provided by the Joint Science Operation Center (JSOC). We are grateful to the RHESSI team for their open data and online quick look service\footnote{http://sprg.ssl.berkeley.edu/~tohban/browser/}. 
We acknowledge the anonymous referee who has helped us in presenting our work in a clearer and more convincing manner. AM is grateful to Kasper Arzner for discussions on various theoretical aspects and for making available his Ph.D. thesis. AM thanks Sasikumar Raja and Rohit Sharma for discussions and suggestions at various stages of the work. The authors thank Culgoora and Learmonth Observatories for making their daily spectrograph data available. 
This work has made use of the NASA ADS service. The authors gratefully acknowledge support from the US Air Force Office of Scientific Research (AFOSR) under Award No. FA9550-14-1-0192. We thank the developers of CASA, SolarSoft Ware, Python 2.7 and the various analysis modules. 



\facility{Murchison Widefield Array, SDO(AIA \& HMI), GOES, RHESSI}

\software{CASA,
SolarSoft Ware,
VCA-NLFFF,
Python 2.7\footnote{https://docs.python.org/2/index.html},
NumPy\footnote{https: //docs.scipy.org/doc/},
Astropy\footnote{http://docs.astropy.org/en/stable/},
Matplotlib\footnote{http://matplotlib.org/}
}
\bibliography{allref}

\appendix

\section{Image registration to a common reference frame} \label{app1} 
As discussed in Section \ref{obs}, the selfcal procedure does not preserve the absolute sky locations of the images, resulting in shifts between images produced using independent selfcal runs. To align the images for a given spectral channel, we used the following procedure.
\begin{itemize}
    \item We chose a time-stamp such that the full solar disk is visible in all images, across the entire frequency band of observation. This was labelled as the reference-time. We found the centroids of the images across frequency for this time. These centroid locations were chosen as the reference location of the Sun as a function of frequency. \textbf{The rest of the procedure was followed independently for each imaging frequency.} 
    \item The calibration tables obtained via selfcal for the reference-time were preserved. For every arbitrary time, we applied the relevant reference-time solutions to the data and imaged it. This yields solar images for the same reference frame as the reference-time. These images were referred to as the true-location images.
    \item We use the fact that all burst time images show the presence of one or more compact sources. We recorded either, the location of the brightest compact source in the image or, the centroid of all compact sources in every true-location image. Later, an independent selfcal was done on the data to arrive at improved images with dynamic range typically close to 1000 at every time slice. 
    \item Once these final images were obtained, we again recorded the location of either, the brightest compact source or the centroid of all compact sources for each of them. The shift in the location of the sources, with respect to the corresponding time-location images, were recorded for each time. These shift estimates for every image were saved into a file. 
    \item The independently self calibrated images were corrected for the shifts in order to bring all the images for a given spectral channel to a common reference frame. 
\end{itemize}

\section{Viability of a double source model for the pulsating burst emission region} \label{app2}
\begin{figure}[ht]
    \centering
    \includegraphics[scale=0.25]{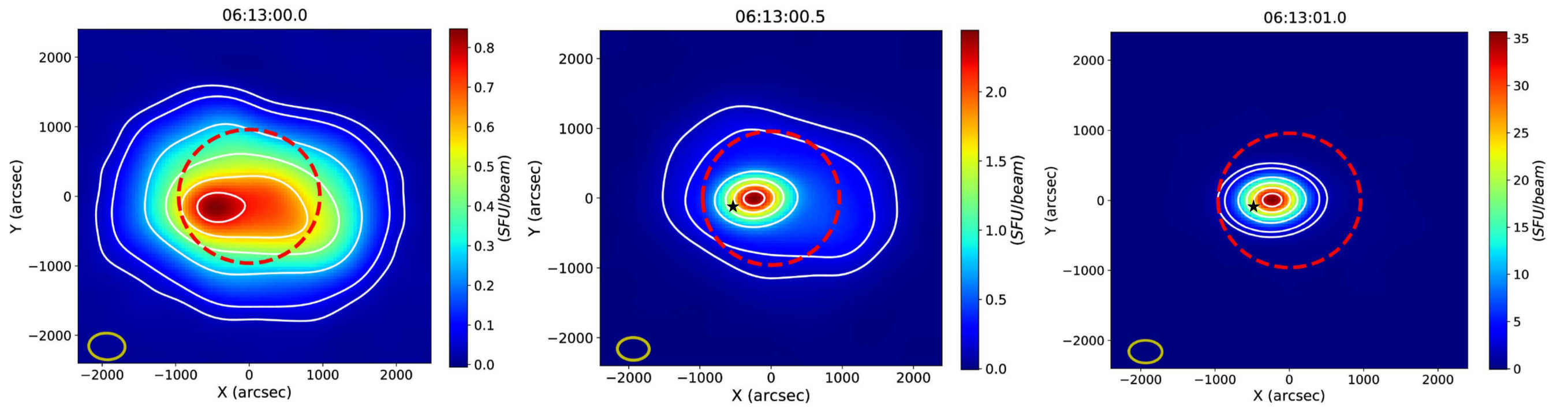}
    \caption{{A montage of three successive solar radio images, each spanning 0.5 s, from close to the start of the first group of type III bursts. 
    The dashed circle marks the optical disk and the ellipse at the bottom left marks the synthesized beam. 
    The left panel comes from a time when the burst emission had not yet picked up, and is typical of the images when the type III burst emission is not present. The entire solar disc is of relatively uniform brightness, varying only by a factor of $\approx$ 2 across the entire visible disc ($\approx$ 0.4 -- 0.8 SFU/beam). The middle panel of Fig. 1 shows the next 0.5 s of data, where a new compact source has appeared at a location about $\approx$ 4.5$^{\prime}$ away from the earlier maximal flux density region with $\approx$ 3 times larger flux density/beam. This is the first radio frame where the source of type III emission is seen. The right panel shows the next frame, where the type III source has maintained its location and its peak flux density/beam has now increased by a factor of $\approx$ 45 over the left panel.
    The very first data point used in Figs. 5 and 6 come from the best fit Gaussian to the source in this frame. Over the next few frames, this compact source continues to steadily brighten till it reaches a peak flux density of about 250 SFU/beam. The black star in the middle and right panels marks the position of peak flux density/beam seen in the left panel, before the type III source had appeared.}}
    \label{qs-act}
    \vspace{-0.2cm}
\end{figure}

\begin{figure}
    \centering
    \includegraphics[height=.48\textwidth,width=0.8\textwidth]{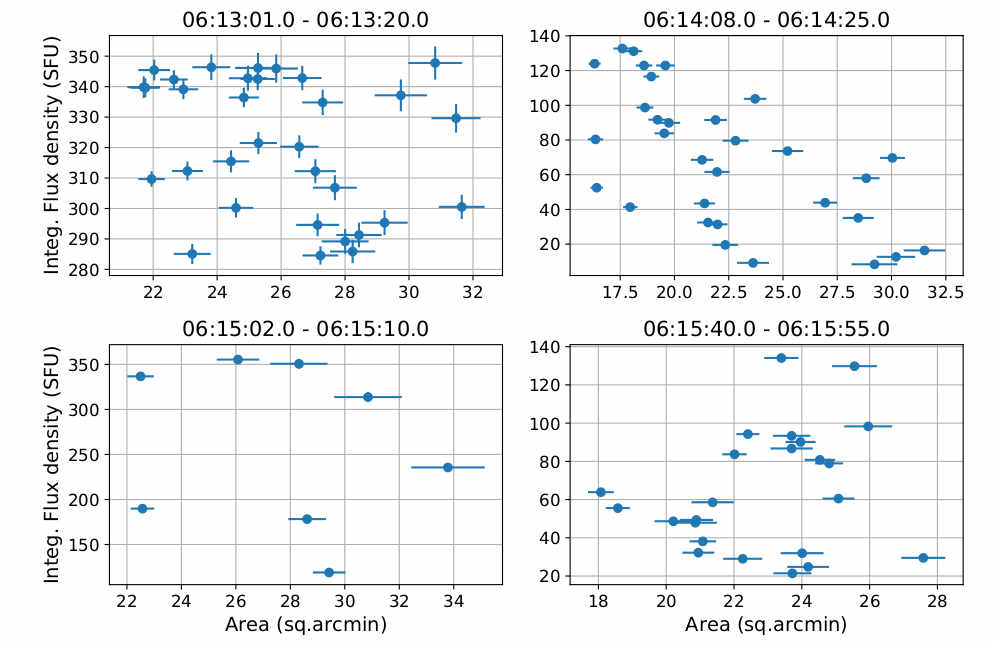}
    \caption{{Integrated flux density as a function of area of the source of burst emission for four groups of bursts in the QPO phase. Only the groups with sufficient data in this phase were chosen. The error bars in the estimates of integrated flux density are typically $< 5\ SFU$ and are hence not visible in most of the panels. There is no clear trend across different burst groups for the source size to become smaller with increasing integrated flux density. The scatter in the points is much larger than the uncertainty associated with them, implying that the scatter is real.}}
    \label{area_integ_allgrps}
    \vspace{-0.2cm}
\end{figure}

{A scenario involving a persistent extended background source and a compact intermitternt variable source which can naturally lead to the observed anti-phased QPOs between the area and integrated flux density of the source of burst emission was presented in Sec. \ref{varsrc}. 
Here we examine this hypothesis in detail.


We start by noting that the imaging dynamic range of the MWA solar radio images typically exceeds 1000, especially when the burst emission is present. An implication is that these images are able to reliably capture emission features significantly weaker than a percent. They are hence of sufficient quality to be able to test the proposed hypothesis and come up with a more detailed model, if necessary.
To examine this possibility in detail we first list some predictions/ consequences of the proposed hypothesis (stated in {\it italics} below) and then test them systematically.
\begin{enumerate}
    \item {\it A persistent extended weak source must exist at or close to the location of burst emission.}\\
    Figure \ref{qs-act} establishes that before the onset of  bursts, no suitable source was present at, or close to, the location where the burst emission would appear. This demonstrates the lack of existence of a background source prior to the start of the groups of bursts. We have examined the images from the intervening periods between successive groups of bursts and conclude the following. The morphology of the quiet Sun images is seen to vary some what and a compact source source is seen to appear occasionally co-located with the hot coronal loop seen in the AIA and RHESSI images (Fig.\ref{mwa+aia} \& \ref{compr}). No evidence for the presence of a weak extended persistent source at or close to the location of the source of type III bursts is seen during these periods. The location of the brightest pixel in the quiet Sun image prior to the onset of bursts was $\approx 4.5'$ away from the centre of the burst source (Fig.\ref{qs-act}). So, for the proposed hypothesis to continue to be valid, the background source must also appear and disappear in tandem with the intermittent compact source.
    \item {\it The area and integrated flux density of the 2D Gaussian model must approach the respective values for the weak extended persistent source at times when the QPOs hit the local minimas in integrated flux density.}\\
\begin{table}
\begin{center}
\begin{tabular}{ |c|c|c|c|c|c|c| } 
 \hline
 Group & I & II & III & IV & V & VI\\ 
 \hline 
 $S_{min}\ (SFU)$&284.56 & 219.88 & 8.36 &23.17 &118.94 &21.41\\
 $S_{max}\ (SFU)$&347.76 & 291.95 & 132.72&31.77 &355.34 &134.1\\
 $Area_{min}$&21.7 & 20.5 & 16.3& 20.7 & 22.5 & 18.1\\
 $Area_{max}$&31.6 & 20.9 & 31.5& 24.3 & 33.8 & 27.6\\
 \hline
\end{tabular}
\end{center}
\caption{The minimum and maximum integrated flux density and area during various burst groups. The values were noted post-linear rise phase so as to cover the QPO phase alone. Compared to the integrated flux density levels, the variation in area is significantly small. Note that these estimates are not co-temporal.}
\label{tab1}
\end{table}
    Table \ref{tab1} lists the minimum and maximum values of the integrated flux density and area of the source of burst emission in the QPO phase (beyond the phase of linear growth in the area). 
    In the QPO phase, the minimum value of observed integrated flux density, $S_{min}$ varies across these groups of bursts from $\approx$ 8 SFU to $\approx$ 285 SFU, while the $Area_{max}$ varies only by about 60\%. 
    For the background source model to remain valid, the background source must vary by up to an order of magnitude on a time scale of order 10 s (e.g. between groups 06:14:52.0 -- 04:14:55.0 UT \& 06:15:02 -- 06:15:10.0 UT) while its area remains essentially unchanged. 
    In other words, not only must the background source appear and disappear with the burst source, its integrated flux density must also change in consonance with it, while maintaining a roughly constant area.
    
    \item {\it An implication of this hypothesis is that the estimated area of the burst source should be tied to the relative strength of the two sources, and should be a smooth and monotonic function of the integrated flux density.
    At the low flux density end, where the relative strength of the compact source is very small, it should be close to the area of the persistent extended source. As the relative strength of the compact intermittent source grows, area should reduce and asymptote to the area of the compact source when its flux density becomes dominant.
    }\\
    The relative variation in the area of the source of burst emission have already been shown to be much smaller than that observed in integrated flux density of the source of burst emission, both within a given group of bursts and across all of them (Table \ref{tab1}).
    Figure \ref{area_integ_allgrps} shows integrated flux density of the burst source as a function of area of the source of burst emission for four of the burst groups which have a significant number of data points in the QPO phase.
    One might argue that the upper envelope of the data in top right panel in Fig. \ref{area_integ_allgrps} follows the trend of decreasing integrated flux density with increasing area predicted by this hypothesis.
    However, at any chosen value of area, there is a very large scatter in the integrated flux density values, much larger than the uncertainty estimates which are too small to be visible on this scale.  
    It is hard to identify any clear trend in these plots.
    
    \item {\it The single 2D Gaussian fits from our analysis should depart the most from the model based on the hypothesis when the source size transitions from being close to compact,  to being close to extended. Hence, for a sufficiently sensitive fitting process, the uncertainty of the fit should be larger in the region when one is transitioning from an extended to a compact source or vice-versa.}\\
    Figure \ref{area_scat_colmap} shows the uncertainty in the estimate of the area of the source of burst emission as a function of the area itself during the QPO phase. Colors denote the integrated flux density and the marker shapes denote the different groups of bursts. 
    We note that the uncertainties in the estimates of area of the source of burst emission lie in the range of a few percent, rising up to $\approx$ 4\% for the weakest instances of burst emission. 
    We also note that bursts with vastly different integrated flux densities tend to have very similar uncertainties in the area of the source of burst emission. 
    This suggests that even for the weaker bursts, the uncertainty of the fit is not dominated by signal-to-noise considerations, but reflect an intrinsic ability of the 2D Gaussian model to describe the image well.
    If the proposed hypothesis were to be true, one would expect there to be a bulge along the Y-axis in the distribution of points on this plot in the region around 20 $sq. arcmin$ -- 25 $sq. arcmin$. 
    Instead, one sees a monotonically increasing trend with larger sources having marginally larger associated uncertainties. 
 \end{enumerate}
 We conclude that if the extended weak source were to indeed exist, it must track multiple properties of the burst source to an implausible degree. 
 None of the predicted trends which were investigated - the relative strength of the emission; variation of size of the source with its integrated flux density; and dependence of uncertainty in the area of the source of burst emission on the area itself were observed.
 This hypothesis can hence be ruled out.
}
\vspace{-1.4cm}
\begin{figure}
    \centering
    \includegraphics[scale=0.5]{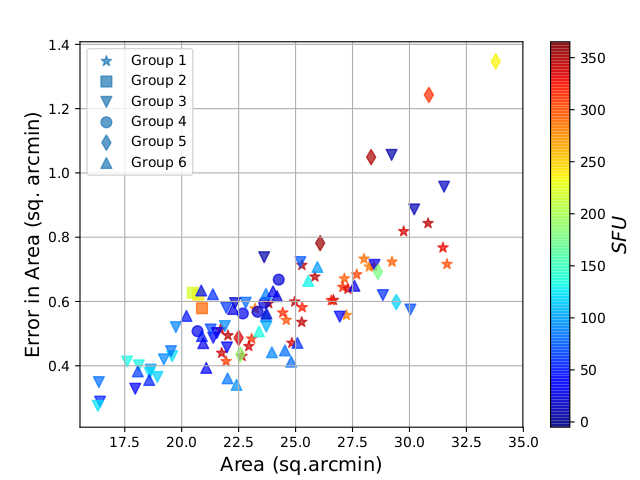}
    \caption{{The area of the source of burst emission against the estimated uncertainty in its estimate for observations across all burst groups during the QPO phase.
    Different markers denote different burst groups and colors denote the integrated flux density.
    }}
    \label{area_scat_colmap}
    \vspace{-0.2cm}
\end{figure}



\end{document}